# Wave propagation through disordered media without backscattering and intensity variations

Konstantinos G. Makris[1], Andre Brandstötter[2], Philipp Ambichl[2],

Ziad H. Musslimani[3], and Stefan Rotter[2]

[1]Crete Center for Quantum Complexity and Nanotechnology, Department of Physics, University of Crete, Heraklion 71003, Greece

[2]Institute for Theoretical Physics, Vienna University of Technology, Vienna 1040, Austria

[3]Department of Mathematics, Florida State University, Tallahassee, Florida 32306, USA

Email: KGM, makris@physics.uoc.gr; AB, andre.brandstoetter@tuwien.ac.at; PA, philipp.ambichl@tuwien.ac.at; ZHM, muslimani@math.fsu.edu; SR, stefan.rotter@tuwien.ac.at

Correspondence: KGM, Email: makris@physics.uoc.gr, Phone: 30-2810-394200

### Abstract

A fundamental manifestation of wave scattering in a disordered medium is the highly complex intensity pattern the waves acquire due to multi-path interference. Here we show that these intensity variations can be entirely suppressed by adding disorder-specific gain and loss components to the medium. The resulting constant-intensity (CI) waves in such non-Hermitian scattering landscapes are free of any backscattering and feature perfect transmission through the disorder. An experimental demonstration of these unique wave states is envisioned based on spatially modulated pump beams that can flexibly control the gain and loss components in an active medium.

**Keywords:** Anderson localization; complex media; non-Hermitian physics; scattering

# INTRODUCTION

The scattering of waves through disordered media is a paradigmatic phenomenon that has captured the interest of various communities for quite some time now[1,2,3]. Among the many important physical aspects of wave propagation that have been studied the phenomenon of Anderson localization has received particular attention[4,5,6,7,8,9,10,11,12]. While much work has been invested into understanding the *statistical* properties of the corresponding wave phenomena[13] there has recently been a surge of interest in controlling the scattering of waves through *individual* systems for specific purposes such as detection, imaging, and efficient transmission across disordered materials[14]. Remarkable progress in these endeavours has recently been made in the optical domain, largely due to the availability of spatial light modulators and new concepts for how to apply them on turbid media[15,16,17]. In a first generation of corresponding experiments the focus was laid on shaping the input wave front impinging on an immutable disordered sample such as to achieve a desired output, like a spatial or temporal focus behind the medium[18,19,20,21]. More recent studies focused instead on controlling the medium itself, e.g., through the material fabrication process[22] or through a spatially modulated pumping[23], leading, e.g., to a versatile control of random and micro-cavity lasers[24,25,26,27,28].

Largely in parallel to these efforts on disordered media, it was recently realised that materials and devices can get entirely new functionalities when adding to them a suitably arranged combination of gain and loss. In particular, structures with a so-called parity-time (PT) symmetry[29,30], were recently introduced theoretically[31,32,33] and experimentally[34,35] in the context of paraxial waveguide optics. Based on a delicate balance between gain and loss, these synthetic structures exhibit rich and unconventional behaviour, holding promise for numerous applications in nano photonics and lasers[36]. In particular, the relation between coherent perfect absorption and scattering through PT-cavities[37] as well as unidirectional invisibility in PT-gratings[38] have attracted a lot of attention. Symmetry breaking in fiber loop optical networks[39], PT-scattering structures[40], and periodic PT-systems as new type of metamaterials[41] are also active research directions in this new field on PT-optics. Along with these activities, another direction is that of complex lasers that rely on the concepts of PT symmetry breaking and exceptional points. Such synthetic lasers with novel characteristics, are based on loss engineering[42]. More specifically, coupled PT-symmetric micro disk lasers can lead to optical diodes[43], single mode microring lasers[44], synthetic PT-lasers[45], loss-induced lasing[46,47] and lasers with chiral modes[48,49]. More recent developments include large scale exceptional points in multilayer optical geometries[50], transient growth in non-normal lossy potentials[51], modulation instabilities in non-Hermitian structures[52], non-Hermitian phase matching in optical parametric oscillators[53], higher order exceptional points[54], protocols for asymmetric mode switching based on encircling exceptional points dynamically[55,56], and directional cloaking based on non-Hermitian potentials[57].

Here we will build on the advances that were made in both of the above research fields with the aim to combine them in a novel and potentially very useful way: We show that for a general disordered medium, given by a distribution of the real part of the refractive index $n_R(x)$, a corresponding distribution of its imaginary part $n_I(x)$ can be found, such that a wave propagating through this medium will feature a constant intensity throughout the entire non-uniform scattering landscape. In other words, we demonstrate that adding a judiciously chosen distribution of gain and loss to a disordered medium will make waves lose all of their internal intensity variations such as to propagate through the disorder without any back reflection†.

---

† see also our conference proceeding: Makris KG, Brandstötter A, Ambichl P, Musslimani ZH, and Rotter S, "Constant intensity waves and transmission through non-Hermitian disordered media" in *Frontiers in Optics 2016*, OSA Technical Digest, paper JTh2A.4. (submitted on 25th of May, 2016)

# MATERIALS AND METHODS

**Scattering states without internal intensity variations.** The solution strategy that we explore for this purpose is based on the one-dimensional Helmholtz equation that describes time-independent scattering of a linearly polarised electric field $\psi(x)$ both in forward and in backward direction,

$$[\partial_x^2 + \varepsilon(x)k^2]\psi(x) = 0. \tag{1}$$

Here $\varepsilon(x)$ is the dielectric permittivity function varying along the spatial coordinate $x$, $k = 2\pi/\lambda$ is the wavenumber (with $\lambda$ being the free space wavelength) and $\partial_x \equiv d/dx$. The dielectric function is complex thus $\psi(x) = [n_R(x) + i n_I(x)]^2$. In general, when a plane wave is incident on a spatially varying distribution $\varepsilon(x)$, interference takes place between the waves propagating forward and backward. As a result, a complex interference pattern is produced with spatial variations on its intensity. As we will now show, this fundamental physical picture can be quite different in the case of non-Hermitian cavities with loss and/or gain.

To jump right to the heart of the matter, we start with an ansatz for a constant-intensity (CI) wave with unit amplitude, $\psi(x) = \exp[iS(x)]$, where $S(x)$ is a real valued function. Due to the obvious relation to WKB-theory[58], we will derive the CI-solution of the Helmholtz Eq. (1) in the bulk, by demanding that the ansatz $\psi(x) = \exp[iS(x)]$ has to be exact in the first order WKB-approximation. As a result of this analysis (see supplemental material for details) we obtain the following non-Hermitian dielectric function,

$$\varepsilon(\mathrm{x}) = W^2(\mathrm{x}) - \frac{i}{k}\partial_x W(x), \qquad (2)$$

that supports a corresponding CI-solution $\psi(x) = \exp[ik\int W(x')dx']$ at wavenumber $k$, which solves Eq. (1) exactly and is valid for the whole bulk space. At this point we have to emphasise, that the above exact WKB analysis is generally valid (not only in the geometric optic limit). The fact that $W(x)$ can be chosen arbitrarily, with no limitations on its spatial complexity (apart from smoothness), is a key asset of this approach, making it very generally applicable. From this result it is also clear that for vanishing imaginary part $[W(x) = const.]$, the dielectric function, as defined in Eq. (2), reduces to $\varepsilon(x) = const.$, resulting in the familiar plane wave in a homogeneous dielectric or in free space. We emphasise that our approach works not only when the functional profile $W(x)$ is known from the outset. Also when a refractive index distribution $n_R(x)$ is given, the corresponding gain-loss-profile $n_I(x)$ can be determined (see supplemental material for details).

Furthermore, it can be shown that CI-waves can also be found for all dielectric functions that are described by Eq. (2) in a finite domain $x \in [-D, D]$, bordering on free space for $x < -D$ and $x > D$. In this case, the scalar Helmholtz-type Eq. (1) admits the following exact CI-scattering-state:

$$\psi(x) = \begin{cases} \exp[ik(x+D)], & x < -D, \\ \exp[ik \int_{-D}^{x} W(x')dx'], & -D \leq x \leq D, \\ \exp[ik(x-D+c)], & x > D, \end{cases} \quad (3)$$

where the integration constant $c$ makes sure that the field continuity relations[59] are satisfied and is given by $c = \int_{-D}^{D} W(x)dx$. We note at this point that the wavenumber $k$ appearing in Eq. (2) is the same as the wavenumber $k$ in the above CI-wave solution. In other words, for any value of $k$ for which a CI-scattering state is desired, the dielectric function $\varepsilon(x)$ has to be engineered correspondingly. Once $\varepsilon(x)$ is fixed and plane waves with varying values of $k$ are impinging on this dielectric structure, a perfectly transmitting CI-solution in general only occurs at the predetermined $k$ value inherent in the design of $\varepsilon(x)$. Due to this restriction to a single frequency also no issues arise with the Kramers-Kronig relations.

Most importantly, the solution in Eq. (3) does not only feature a constant intensity $|\psi(x)|^2 = 1$ in the asymptotic regions $x \leq -D$ and $x \geq D$, where $\varepsilon(x) = 1$ and simple plane wave propagation is realised, but also inside the finite region of length $2D$ in which the dielectric function varies and the phase-evolution is non-trivial. Regarding the appropriate boundary conditions at $x = \pm D$, it can be shown that the following perfect transmission boundary conditions (zero reflection)[40],

$$\partial_x \psi(\pm D) = ik\psi(\pm D) \qquad (4)$$

imply the following conditions for the generating function,

$$W(D) = 1 = W(-D). \qquad (5)$$

**Non-Hermitian scattering methods.** In order to elucidate the above ideas, we consider in the "Results and Discussions" section several specific examples of index distributions and study the CI-waves they give rise to. For these calculations, a transfer matrix method[59] was used for TE-linearly polarised optical waves along with an effective Hamiltonian approach[17]. More specifically, the transfer matrix method is valid for piecewise refractive index distributions. In order to apply such an approach to our scattering problem, we discretised the continuous potential into many slabs of almost constant refractive index, and then applied the transfer matrix method. Another technique we used was that of the effective Hamiltonian. The wave equation Eq. (1), $[\partial_x^2 + \varepsilon(x)k^2]\psi(x) = 0$, can be written as a generalised eigenvalue problem for the potential of Eq. (2) for a given $W(x)$. More specifically we have: $\partial_x^2 \psi(x) = -\varepsilon(x,k)\tilde{k}^2 \psi(x)$. This generalised eigenvalue problem is non-Hermitian due to the perfect transmission boundary conditions. Notice that only one of its eigenmodes will be the constant-intensity state and satisfy the relation $\tilde{k} = k$. We have compared the two different approaches for specific optical structures and they give identical results.

**Materials.** A crucial question is the physical values of gain and loss required to observe and realise the proposed constant-intensity waves. It turns out that the required gain and loss values depend directly on the slope and amplitude of the refractive index distribution, as well as on the wavelength of operation, as we can see from Eq. (2). By delicately choosing those values, realistic values for gain and loss can be achieved with existing technology (actual gain and loss values for the simulations shown are provided in the next section).

## RESULTS AND DISCUSSIONS

As a first example, we assume $W(x)$ to be a parabolic function modulated with a cosine, namely $W(x) = [1 - 0.2\cos(15\pi x / 2)]\exp(2 - x^2)$. The corresponding real part of the refractive index distribution $n_R(x)$ is shown as the grey shaded area in Fig. 1. A wave impinging on the dielectric structure composed of only $n_R(x)$ is partly reflected and features a highly oscillatory profile, see Fig. 1(a). Quite in contrast, when adding also the gain and loss inherent in the imaginary index component $n_I(x)$ derived from $W(x)$ [see green and red regions in Fig. 1(b)], the resulting scattering state is fully transmitted and features a constant intensity. Because of the boundary conditions, $W(x)$ must be symmetric at the endpoints of the cavity, resulting in an anti-symmetric distribution of $n_I(x)$. Our example shows that for a plane wave at an arbitrary incident wavenumber $k$, we can find the corresponding gain-loss landscape [from Eq. (2)] such that this wave will fully penetrate the scattering medium without forming any spatial variations in its intensity pattern.

**Perfect transmission through disorder.** The most striking application of CI-waves is realised for the case of disordered environments, which is also the focus of our work. We know, e.g., that in strongly scattering disordered media Anderson localization occurs, resulting in an exponential decrease of the transmittance $T = |t|^2$ for structures with sizes greater than the localization length $\xi = -2D \langle \ln[T(D)] \rangle^{-1}$. For a given real and disordered index of refraction in the localised regime close to unit transmittance is thus very unlikely and occurs only at well-isolated, sharply resonant wavenumbers that are difficult to achieve experimentally[60,61]. Our approach now allows to turn this behaviour upside down – not only in the sense that we can engineer unit transmission at any predetermined value of the wavenumber $k$, but also that we can create scattering states that have constant intensity in a strongly disordered environment which would usually give rise to the most dramatic intensity fluctuations known in wave physics.

We illustrate our results for the disordered one-dimensional slab shown in Fig. 2, where a refractive index distribution following Eq. (2) is considered with a tunable imaginary component, $\varepsilon(x) = [n_R(x) + ian_I(x)]^2$ (the tunable parameter $a$ controls the overall amplitude of gain and loss). From Eq. (2), one can understand that CI-scattering-states exist only for $a = 1$. More specifically, we choose the generating function $W(x)$ to be a superposition of $N$

Gaussian functions of the same width $d$, but centered around random positions $c_n$ and having random amplitudes $r_n$. More specifically, we consider $W(x) = \sum_{n=1}^{N} r_n \exp\left[-\left(\frac{x-c_n}{d}\right)^2\right]$, where $c_n$ is a random variable. This leads us to the following analytical expression of a CI-state $\psi(x)$ in a disordered medium:

$$\psi(x) = \begin{cases} \exp[ik(x+D)], & x < -D, \\ \exp\left[\frac{ikd\sqrt{\pi}}{2}\sum_{n=1}^{N} r_n \left\{erf\left(\frac{x-c_n}{d}\right) + erf\left(\frac{D+c_n}{d}\right)\right\}\right], & -D \leq x \leq D, \\ \exp[ik(x-D+c)], & x > D, \end{cases} \quad (6)$$

where the error function is defined as follows: $erf(x) = \frac{2}{\sqrt{\pi}} \int_0^x \exp(-s^2)ds$, and the constant $c$ is defined as in Eq. (3). For $a = 0$ the refractive index is Hermitian, whereas for $a = 1$ CI-waves exist. The refractive index distribution of such a non-Hermitian disordered medium is depicted in Fig. 2(a), and the localization length $\xi$ of the Hermitian refractive index ($a = 0$) is depicted in Fig. 2(b). The imaginary part (gain and loss) of the refractive index distribution that leads [based on Eq. (2)] to a CI-state is depicted in Fig. 2(c). As is physically expected, without the gain and loss distribution, the system reflects almost all waves due to localization. Furthermore, adding first only the gain part of the refractive index distribution that leads to a CI-state [see Fig. 2(c)] still results in highly oscillatory scattering wave functions with finite reflectance for all values of the gain-loss amplitude $a$ (from 0 to 1), as is illustrated in Fig. 2(d). Quite counterintuitively, adding also the loss part of the refractive index distribution leads to perfect transmission without any intensity variations for $a = 1$, as demonstrated in Fig. 2(e). By varying the gain-loss amplitude $a$, as in Fig. 2(e), we can thus see a smooth transition from the Anderson localization regime (at $a=0$) to perfect transmission with constant intensity (at $a=1$). In other words, the presence of gain and loss in a scattering environment can completely suppress the localization of waves due to multiple scattering, leading to a delocalised state with a constant intensity.

The maximum value of gain/loss for the system of Fig. 2 is at $\max(n_I) \approx 6.7 \cdot 10^{-3}$ with a cavity length of 16cm. If we assume a wavelength of $\lambda = 1.6$ μm this would correspond to gain of approximately 526 cm$^{-1}$. Such realistic values of gain should be implementable in a practical experimental setting using, e.g., dye rhodamine (6G) solution materials, typical in active plasmonics[62], and in optofluidic random lasers[63]. Smaller values of gain and loss can be realised when considering smoother refractive index distributions, $n_R(x)$, for which, in turn, the localization length $\xi$ is larger.

**Discrete disordered systems.** Another important aspect of CI-waves is their experimental realization, with the most challenging part being the fabrication of a specific index distribution with gain and loss[64]. In order to overcome such inherent difficulties, we study here also the existence of CI-waves in a system of discrete scatterers, like the one that is presented in Fig. 3. Such a set-up is composed of many discrete elements (cavities) with gain or loss and a specific real refractive index distribution. The analytic solution of Eq. (2) is still valid in the discrete version of the Helmholtz-type wave equation:

$$\varepsilon_m = b^{-2}\left\{2 - e^{\frac{ik\Delta x}{2}(W_m+W_{m+1})} - e^{-\frac{ik\Delta x}{2}(W_m+W_{m-1})}\right\}$$

(7)

and

$$\psi_m = \exp\left[\frac{ik\Delta x}{2}(W_1 + W_M + 2\sum_{n=1}^{M}W_n)\right]$$

(8)

where $\varepsilon_m$ is the permittivity of the $m$-th scatterer, $b = \omega\Delta x$, $\omega^2 = 2[1-\cos(k\Delta x)]/\Delta x^2$ and $m = 1,...,M$. Additionally, the perfect transmission boundary conditions imposed at the endpoints of the discrete chain of scatterers $\psi_0 = \psi_1\exp(-ik\Delta x)$, and $\psi_{M+1} = \psi_M\exp(ik\Delta x)$ as well as the relation $\omega\Delta x < 2$ must always hold. We consider a specific example in Fig. 3 of $M$ elements that form a one-dimensional disordered chain. By adding gain or loss onto the sites as prescribed by Eq. (7) an incoming wave from the left will have the same constant intensity on all of these sites.

**PT symmetry and mean reality condition.** So far we haven't discussed the relation of the non-Hermitian distribution [Eq. (2)] that supports CI-waves with PT symmetry. For the special case that the generating function is even with respect to $x$, namely $W(x) = W(-x)$, the dielectric function turns out to be PT-symmetric since $\varepsilon(x) = \varepsilon^*(-x)$. In other words, our approach is rather general and the only restrictions are the permittivity distribution [Eq. (2)] and the boundary conditions [Eq. (5)] for $W$. Keep in mind that $\text{Re}[\varepsilon(x)]$ could in principle also be negative - at least there is no restriction from the mathematical point of view. Since we wish to study relevant physical materials that are easily accessible also experimentally, we choose our $W(x)$ such that we have $\text{Re}[\varepsilon(x)] > 1$ and also $n_R(x) > 1$. A direct consequence of these two restrictions is that the spatial average gain-loss over the cavity region is zero, $\int_{-D}^{D} \text{Im}[\varepsilon(x)] dx = 0$. In terms of the refractive index, the evaluation of the integral $\int_{-D}^{D} n_I(x) dx$ depends on the symmetry of $W(x)$. For example, if $W(x)$ is an even function of $x$, then the refractive index is PT-symmetric, and $\int_{-D}^{D} n_I(x) dx = 0$. It would be interesting to explore if this condition of "mean reality" also has other interesting consequences that go beyond the restrictions imposed by PT symmetry.

**Effect of wavenumber detuning and gain saturation.** As we have already shown analytically and numerically, for a fixed refractive index determined by Eq. (2), a CI-wave at the corresponding wavenumber $k_0$ exists. A natural question one may ask at this point is what happens to incident plane waves with detuned wavenumbers $k \neq k_0$ (Fig. 4) (when considering active materials that are characterised by approximately flat dispersion curves near the values of the wavelength of operation). Naively, one may expect that the emergence of CI-waves is a sharp resonance phenomenon, so that waves with a slight detuning in the wavenumber $k_0$ should show a completely different behaviour as, e.g., around a resonance in a Fabry-Pérot interferometer[59]. This picture turns out to be misleading on several levels: Since the CI-wave function at position $x$, $\psi(x) = \exp[ik \int_{-D}^{x} W(x')dx']$, only depends on the generating function $W(x')$ evaluated at values $x' < x$, one can easily truncate the system at any point $x$ and still get a CI-wave – provided one continues the system for all $x' > x$ with a constant generating function that has the same value as at the point of truncation. This behaviour indicates that a refractive index profile that supports CI-waves is not only reflectionless in total, but also unidirectional for the given structure. Perfect transmission in such systems is thus not a resonance phenomenon (as resulting from a back and forth propagation of waves), suggesting that CI-waves are stable against changes of the incident wavelength. To check this explicitly, we numerically calculated the average resonance width of the transmission spectrum $|t(k)|$ of the Hermitian system in Fig. 1, $\langle \Delta k_{Hern} \rangle = 0.74$, in an interval $k \in [\frac{2\pi}{0.26} - 3, \frac{2\pi}{0.26} + 3]$, with minimum transmission $|t(k)_{min}| = 0.94$, as is shown in Fig. 4. The transmission of waves through the corresponding non-Hermitian CI-refractive index [that of Fig. 1(b)] stays larger than 0.99 over the entire k-interval, confirming our prediction. Another important point to make is that one can easily achieve a transmission equal to one in a non-Hermitian system just by adding enough gain to it. In the scattering landscapes that we consider here, however, the net average amplification is zero, since $\int_{-D}^{D} \text{Im}[\varepsilon(x)]dx = 0$ and the intensity is equally distributed everywhere.

In realistic gain materials, high enough field amplitudes lead inevitably to gain saturation. A natural question to ask is if CI-waves exist also in this nonlinear regime. As we show in the supplemental material, one can analytically derive CI-scattering-states by solving the corresponding nonlinear Helmholtz equation. The difference in this case is that the CI amplitude is specific for the parameters of the gain material.

**Connections to the literature.** The results presented above also have several interesting connections to earlier insights from the literature. Perhaps the first mention of CI-waves was made in laser physics in an under-appreciated work by Yariv *et al.*[65], in which it was shown that modes in distributed feedback (DFB) lasers can be engineered to have a constant intensity throughout the entire laser device – a feature that was proposed as a strategy to overcome spatial and spectral lasing instabilities[66,67]. The design principle for CI-waves in these DFB lasers was, however, restricted to periodic potential structures without any incoming wave and is thus orthogonal to our own approach. In a recent work, CI-waves were presented as solutions of the paraxial wave equation where the "Wadati potential", previously introduced in Ref. [68], varies only transversely to the propagation direction[52]. The paraxial approximation employed in that work, however, excludes any backscattering from the potential and the incoming wave had to be engineered through wave front shaping to yield the desired CI-solution. The approach presented here has the clear advantage that it does not rely on any approximation, that no wave front shaping of the incoming wave is necessary and that it can be applied to arbitrary, even disordered potentials with an unspecified amount of backscattering. We also mention in this context that during the last few years non-Hermitian potentials without PT symmetry that yield real propagation constants [including similar ones as in our Eq. (2)] have been studied (see Refs. [69-73]). In our own work we are, however, not concerned with a phase transition to complex eigenvalues, but rather focus on the unique possibility to achieve perfect transmission and a suppression of any intensity variations in disordered media. For this purpose it is clearly essential that, in contrast to these earlier works, we address here the full scattering problem including backscattering. Last but not least we also highlight that our approach opens up a new and promising way to apply the WKB approximation to potentials such as those with a short-range disorder, that usually fall completely outside the scope of this well-studied approximation.

## CONCLUSIONS

In conclusion, we examine the existence and the properties of a novel type of waves, the constant-intensity (CI) waves in one-dimensional non-Hermitian optical slab geometries with and without disorder. For any wavenumber $k$ of a plane wave incident on a real refractive index distribution, we can identify a corresponding gain-loss profile that allows for CI-waves to propagate through such a scattering landscape without any reflection. Most importantly, we found the gain-loss profiles that need to be added to any disordered system such as to completely overcome the strong backscattering and the intensity variations that usually occur in disordered media. As a first step towards an experimental realization we propose to study chains of discrete scatterers with gain and loss that can nowadays be routinely fabricated in the laboratory.

## ACKNOWLEDGMENTS

This project was supported by the People Programme (Marie Curie Actions) of the European Union's Seventh Framework Programme (FP7/2007-2013) under REA Grant Agreement No. PIOFGA-2011-303228 (Project NOLACOME). K.G.M. is also supported by the European Union Seventh Framework Programme (FP7-REGPOT-2012-2013-1) under grant agreement 316165. Z.H.M. was supported in part by NSF Grant No. DMS-0908599. S.R. acknowledges financial support by the Austrian Science Fund (FWF) through Project SFB NextLite (F49-P10) and Project GePartWave (I1142).

KGM and SR developed the concept of the study. KGM, AB, PA, and SR conducted the analysis, data interpretation and drafted the manuscript. ZHM contributed to the development of the numerical and analytical methods, data interpretation and drafting of the manuscript.

## REFERENCES

[1]    Lagendijk A, and Tiggelen BA (1996) Resonant multiple scattering of light, Physics Reports **270**, 143.

[2]    Akkermans E, and Montambaux G (2007) *Mesoscopic Physics of Electrons and Photons*, Cambridge University Press, 1st edition.

[3]     Sebbah P (1999) *Waves and Imaging through complex media*, Kluwer Academic.

[4]     Anderson PW (1958) Absence of diffusion in certain random lattices. Phys. Rev. **109**, 1492.

[5]     Wiersma DS, Bartolini P, Lagendijk A, and Righini R (1997) Localization of light in a disordered medium. Nature **390**, 671.

[6]     Chabanov AA, Stoytchev M, and Genack AZ (2000) Statistical signatures of photon localization. Nature **404**, 850.

[7]     Schwartz T, Bartal G, Fishman S, and Segev M (2007) Transport and Anderson localization in disordered two-dimensional photonic lattices. Nature **446**, 52.

[8]     Lahini Y et al (2008) Anderson localization and nonlinearity in one-dimensional disordered photonic lattices. Phys. Rev. Lett. **100**, 013901.

[9]     Billy J et al. (2008) Direct observation of Anderson localization of matter waves in a controlled disorder. Nature **453**, 891.

[10]    Roati G (2008) et al. Anderson localization of a non-interacting Bose-Einstein condensate. Nature **453**, 895.

[11]    Lagendijk A, Tiggelen B, and Wiersma DS (2009) Fifty years of Anderson localization. Physics Today **62**, 24.

[12]    Segev M, Silberberg Y, and Christodoulides DN (2013) Anderson localization of light. Nat. Phot. **7**, 197.

[13]    Beenakker CWJ (1997) *Random-matrix theory of quantum transport*, Phys. Rep. **69**, 731.

[14] Mosk AP, Lagendijk A, Lerosey G, and Fink M (2012) Controlling waves in space and time for imaging and focusing in complex media. Nat. Phot. **6**, 283.

[15] Vellekoop IM, and Mosk AP (2007) Focusing coherent light through opaque strongly scattering media. Opt. Lett. **32**, 2309.

[16] Popoff SM, Lerosey G, Carminati R, Fink M, Boccara AC, and Gigan S (2010) Measuring the transmission matrix in optics: an approach to the study and control of light propagation in disordered media. Phys. Rev. Lett. **104**, 100601.

[17] Rotter S, Ambichl P, and Libisch F (2011) Generating particlelike scattering states in wave transport. Phys. Rev. Lett. **106**, 120602.

[18] Vellekoop IM, Lagendijk A, and Mosk AP (2010) Exploiting disorder for perfect focusing. Nat. Phot. **4**, 320.

[19] Katz O, Small E, Bromberg Y, and Silberberg Y (2011) Focusing and compression of ultrashort pulses through scattering media. Nat. Phot. **5**, 372.

[20] McCabe DJ, Tajalli A, Austin DR, Bondareff P, Walmsley IA, Gigan S, and Chatel B (2011) Spatio-temporal focusing of an ultrafast pulse through a multiply scattering medium. Nat. Comm. **2**, 447.

[21] Yaqoob Z, Psaltis D, Feld MS, and Yang C (2008) Optical phase conjugation for turbidity suppression in biological samples. Nat. Photonics **2**, 110.

[22] Riboli F et al. (2014) Engineering of light confinement in strongly scattering disordered media. Nat. Mater. **13**, 720.

[23] Bruck R et. al (2016) An all-optical spatial light modulator for field-programmable silicon photonic circuits. arXiv:1601.06679.

[24] Rotter S (2014) Playing pinball with light. Nat. Phys. **10**, 412.

[25] Bachelard N, Gigan S, Noblin X, and Sebbah P (2014) Adaptive pumping for spectral control of random lasers. Nat. Phys. **10**, 426.

[26] Hisch T, Liertzer M, Pogany D, Mintert F, and Rotter S (2013) Pump-controlled directional light emission from random lasers. Phys. Rev. Lett. **111**, 023902.

[27] Ge L, Malik O, and Türeci HE (2014) Enhancement of laser power-efficiency by control of spatial hole burning interactions. Nat. Photonics **8**, 871.

[28] Liew SF, Redding B, Ge L, Solomon GS, and Cao H (2014) Active control of emission directionality of semiconductor microdisk lasers. Appl. Phys. Lett. **104**, 231108.

[29] Bender CM, and Boettcher S (1998) Real Spectra in Non-Hermitian Hamiltonians Having *PT* Symmetry. Phys. Rev. Lett. **80**, 5243.

[30] Bender CM, Brody DC, and Jones HF (2002) Complex Extension of Quantum Mechanics. Phys. Rev. Lett. **89**, 270401.

[31] Makris KG, El-Ganainy R, Christodoulides DN, and Musslimani ZH (2008) Beam Dynamics in *PT* Symmetric Optical Lattices. Phys. Rev. Lett. **100**, 103904.

[32] El-Ganainy R, Makris KG, Christodoulides DN, and Musslimani ZH (2007) Theory of Coupled Optical *PT*-Symmetric Structures. Opt. Lett. **32**, 2632.

[33] Musslimani ZH, Makris KG, El-Ganainy R, and Christodoulides DN (2008) Optical Solitons in *PT* Periodic Potentials. Phys. Rev. Lett. **100**, 030402.

[34] Guo A, Salamo GJ, Duchesne D, Morandotti R, Volatier-Ravat M, Aimez V, Siviloglou GA, and Christodoulides DN (2009) Observation of PT-symmetry breaking in complex optical potentials. Phys. Rev. Lett. **103**, 093902.

[35] Ruter CE, Makris KG, El-Ganainy R, Christodoulides DN, Segev M, and Kip D (2010) Observation of parity-time symmetry in optics. Nature Phys. **6**, 192.

[36] Kottos T (2010) Optical Physics: Broken Symmetry Makes light Work. Nat. Phys. **6**, 166.

[37] Chong YD, Ge L, and Stone AD (2011) *PT*-Symmetry Breaking and Laser-Absorber Modes in Optical Scattering systems. Phys. Rev. Lett. **106**, 093902.

[38] Lin Z, Ramezani H, Eichelkraut T, Kottos T, Cao H, and Christodoulides DN (2011) Unidirectional invisibility induced by PT-symmetric periodic structures. Phys. Rev. Lett. **106**, 213901.

[39] Regensburger A, Bersch C, Miri MA, Onishchukov G, Christodoulides DN, and Peschel U (2012) Parity-time synthetic photonic lattices. Nature **488**, 167.

[40] Ambichl P, Makris KG, Ge L, Chong YD, Stone AD, and Rotter S (2013) Breaking of *PT* Symmetry in Bounded and Unbounded Scattering Systems. Phys. Rev. X **3**, 041030.

[41] Feng L, Xu YL, Fegadolli WS, Lu MH, Oliveira JEB, Almeida VR, Chen YF, and Scherer A (2013) Experimental demonstration of a unidirectional reflectionless parity-time metamaterial at optical frequencies. Nat. Mater. **12**.

[42] Liertzer M, Ge L, Cerjan A, Stone AD, Türeci HE, and Rotter S (2012) Pump-Induced Exceptional Points in Lasers. Phys. Rev. Lett. **108**, 173901.

[43] Peng B, Özdemir SK, Lei F, Monifi F, Gianfreda M, Long GL, Fan S, Nori F, Bender CM, and Yang L (2014) Parity-time-symmetric whispering-gallery microcavities. Nat. Phys. **10**, 394.

[44] Hodaei H, Miri MA, Heinrich M, Christodoulides DN, and Khajavikhan M (2014) Parity-time-symmetric microring lasers. Science **346**, 975.

[45] Feng L, Jing Wong Z, Ma RM, Wang Y, and Zhang X (2014) Single-mode laser by parity-time symmetry breaking. Science **346**, 972.

[46] Peng B, Özdemir SK, Rotter S, Yilmaz H, Liertzer M, Monifi F, Bender CM, Nori F, and Yang L (2014) Loss-induced suppression and revival of lasing. Science **346**, 328.

[47] Brandstetter M, Liertzer M, Deutsch C, Klang P, Schöberl J, Türeci HE, Strasser G, Unterrainer K, and Rotter S (2014) Reversing the pump dependence of a laser at an exceptional point. Nat. Comm. **5**, 4034.

[48] Peng B, Özdemir S, Liertzer M, Chena W, Kramer J, Yilmaz H, Wiersig J, Rotter S, and Yang L (2016) Chiral modes and directional lasing at exceptional points. PNAS - Proc. Natl. Acad. Sci. U.S.A. **113**, 6845.

[49] Miao P, Zhang Z, Sun J, Walasik W, Longhi S, Litchinitser N, and Feng L (2016) Orbital angular momentum microlaser. Science **353**, 464.

[50] Feng L, Zhu X, Yang S, Zhu H, Zhang P, Yin X, Wang Y, and Zhang X (2014) Demonstration of a large-scale optical exceptional point structure. Opt. Express **22** 1760.

[51] Makris KG, Ge L, and Türeci HE (2014) Anomalous transient amplification of waves in non-normal photonic media. Phys. Rev. X. **4**, 041044.

[52] Makris KG, Musslimani ZH, Christodoulides DN, and Rotter S (2015) Constant-intensity waves and their modulation instability in non-Hermitian potentials. Nat. Commun. **6**, 7257.

[53] El-Ganainy R, Dadap JI, and Osgood Jr RM (2015) Optical parametric amplification via non-Hermitian phase matching. Opt. Lett. **40**, 5086.

[54] Demange G, and Graefe EM (2012) Signatures of three coalescing eigenfunctions. J. Phys. A: Math. Theor. **45**, 025303.

[55] Doppler J, Mailybaev A, Böhm J, Kuhl U, Girschik A, Libisch F, Milburn T, Rabl P, Moiseyev N, and Rotter S (2016) Dynamically encircling an exceptional point for asymmetric mode switching. Nature **537**, 76.

[56] Xu H, Mason D, Jiang L, and Harris J. G. E (2016) Topological energy transfer in an optomechanical system with exceptional points. Nature **537**, 80.

[57] Gbur G (2015) Designing directional cloaks from localized fields, Opt. Lett. **40**, 986.

[58] Bender CM, and Orszag S (1999) *Advanced mathematical methods for scientists and engineers*, Springer, 1st edition.

[59] Yeh P (2005) *Optical waves in layered media*, Wiley-Interscience.

[60] Wang J, and Genack AZ (2011) Transport through modes in random media. Nature **471**, 345.

[61] Pena A, Girschik A, Libisch F, Rotter S, and Chabanov AA (2014) The single-channel regime of transport through random media. Nat. Comm. **5**, 3488.

[62] Berini P, and De Leon I (2012) Surface plasmon-polariton amplifiers and lasers. Nat. Phot. **6**, 16.

[63] Bachelard N, Gigan S, Noblin X, and Sebbah P (2014) Adaptive pumping for spectral control of random lasers. Nat. Phys. **10**, 426.

[64] Szameit A, and Nolte S (2010) Discrete optics in femtosecond-laser-written photonic structures. J. Phys. B: At. Mol. Opt. Phys. **43**, 163001.

[65] Schrans T, and Yariv A (1990) Semiconductor lasers with uniform longitudinal intensity distribution. Appl. Phys. Lett. **56**, 1526.

[66] Kasraian M, and Botez D (1995) Single-lobed far-field radiation pattern from surface-emitting complex-coupled distributed-feedbak diode lasers. Appl. Phys. Lett. **67**, 2783.

[67] Carlson NW, Liew SK, Amantea R, Bour DP, Evans G, and Vangieson EA (1991) Mode discrimination in distributed feedback grating surface emitting lasers containing a buried second-order grating. IEEE J. Quantum Electron. **27**, 1752.

[68] Wadati M (2008) Construction of Parity-Time Symmetric Potential through the Soliton Theory. J. Phys. Soc. Jpn **77**, 074005.

[69] Tsoy EN, Allayarov IM, and Abdullaev FK (2014) Stable localized modes in asymmetric waveguides with gain and loss. Opt. Lett. **39**, 4215-4218.

[70] Nixon S, Yang J (2016) All-real spectra in optical systems with arbitrary gain-and-loss distributions. Phys. Rev. A **93**, 031802.

[71] Miri MA, Heinrich M, and Christodoulides DN (2013) Supersymmetry-generated complex optical potentials with real spectra. Phys. Rev. A **87**, 043819.

[72] Yang J (2014) Symmetry breaking of solitons in one-dimensional parity-time-symmetric optical potentials. Opt. Lett. **39**, 5547-5550.

[73] Konotop VV, and Zezyulin DA (2014) Families of stationary modes in complex potentials. Opt. Lett. **39**, 5535-5538.

[74] Premaratne M, and Agrawal G. P (2011) *Light Propagation in Gain Media: Optical Amplifiers*, Cambridge University Press.

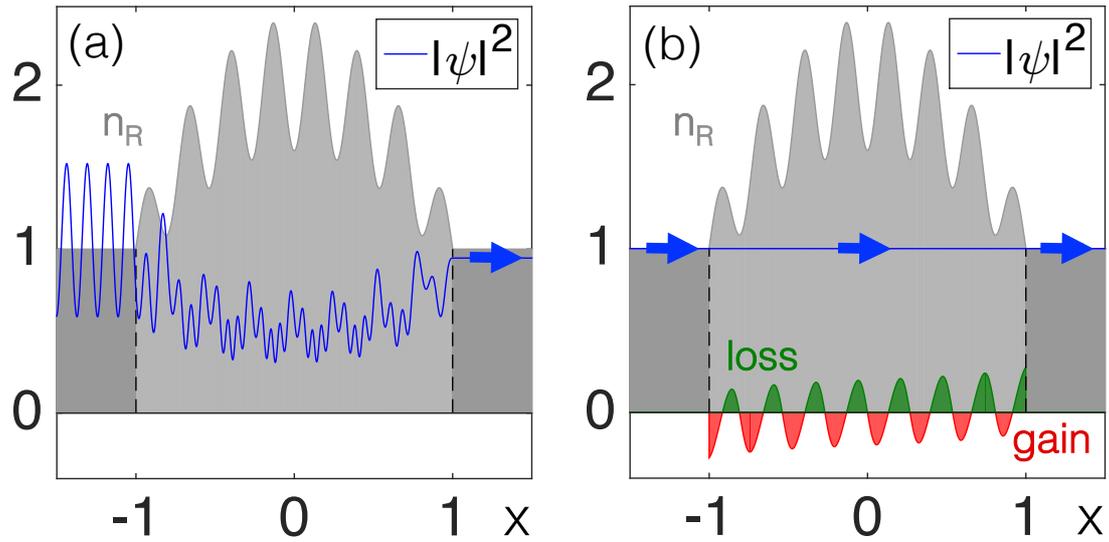

Figure 1: (a) Scattering wave function intensity (blue line) in a Hermitian refractive index for an incident plane wave (from the left) with a specific normalised wavenumber $k = 2\pi / 0.26 = 24.15$. (b) Intensity of the CI-wave for the corresponding non-Hermitian refractive index $n(x)$ and the same incident plane wave. The real part of the refractive index is shown in grey, whereas its imaginary part is coloured in green (loss) and red (gain). For illustration purposes the imaginary part in (b) was multiplied by a factor of 2. The calculations were performed using the transfer matrix approach.

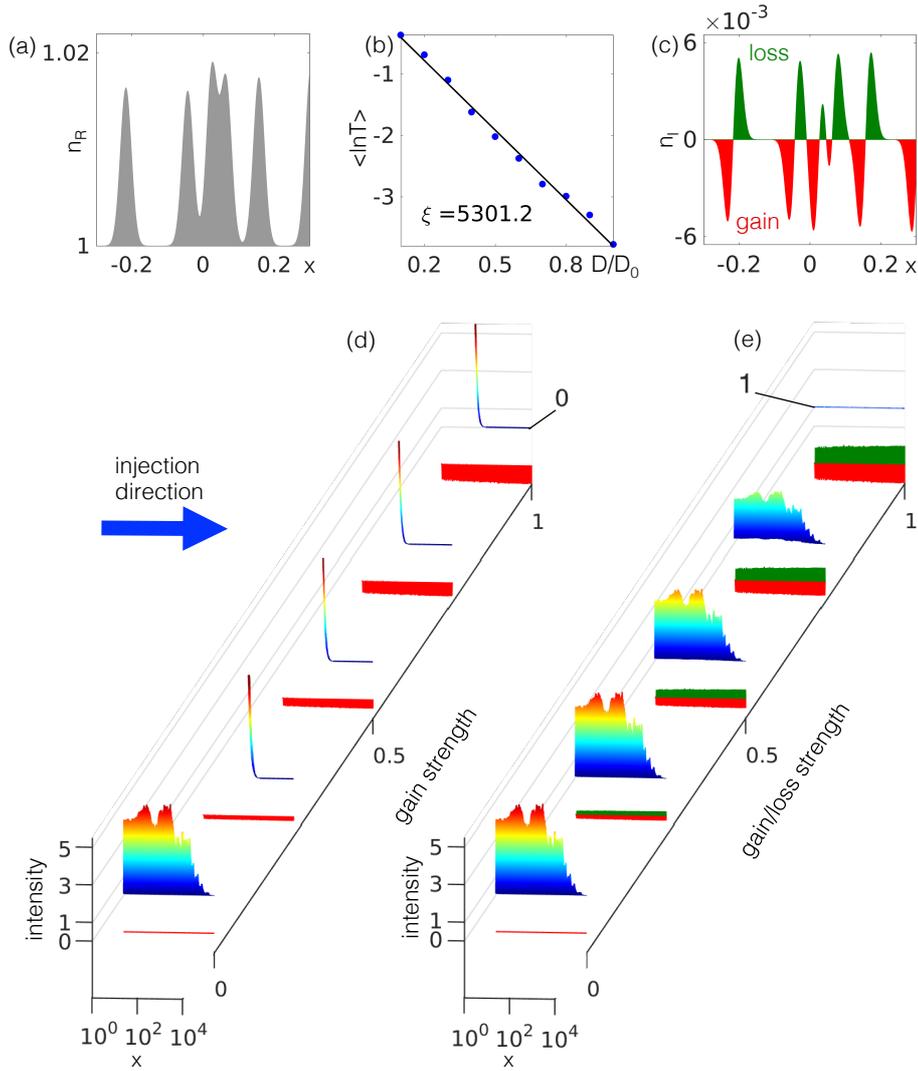

Figure 2: (a) Disordered Hermitian refractive index distribution $n_R(x)$ for $N = 99000$. (b) Exponential suppression of the transmittance $T$ with localization length $\xi$ in the system shown in (a) versus the length cavity $D$ for $D_0 = 5000$. (c) Imaginary part of the refractive index $n_I(x)$ following from the CI design principle for the real index distribution in (a). (d,e) Scattering wave functions for the disordered cavity as a function of the gain strength $a$, for the gain-only and gain-loss potential, respectively. In both cases an incident plane wave with normalised wavenumber $k = 2\pi/0.1 = 62.8$ is considered (from left to right), and $x$-axis is represented in logarithmic scale. The CI-wave can be clearly seen for the full gain-loss strength ($a = 1$) in (e).

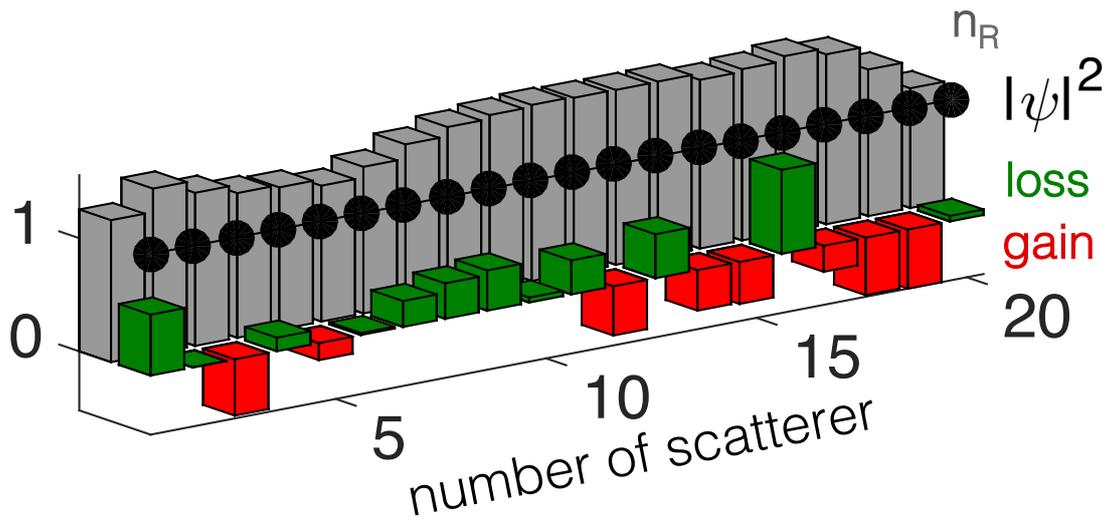

Figure 3: Disordered chain of discrete scatterers with an incoming plane wave from the left. The real part (grey) as well as the gain (red) and loss (green) components of the refractive index are shown for each scatterer. The corresponding discrete CI-wave is depicted with black dots. The normalised parameters used are $M = 20$, $\omega = 12$, $L = 2$ and $\Delta x = L/(M-1)$.

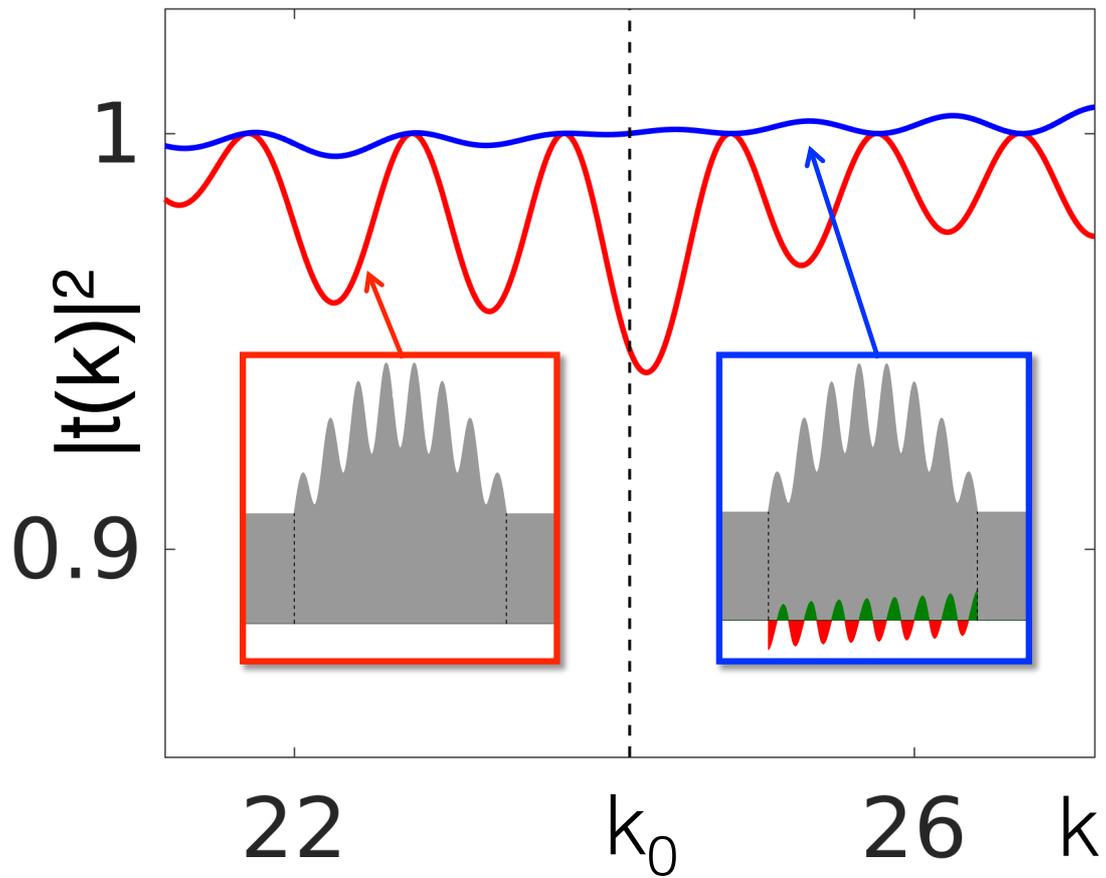

Figure 4: Effect of incident wavenumber detuning in a narrow wavelength window, on the transmittance through the potential of Fig. 1(b) (blue line). The Hermitian case is plotted for comparison (red line). The two insets illustrate the complex refractive index distributions.

# Supplemental material for

# "Wave propagation through disordered media without backscattering and intensity variations"

**WKB analysis.** Expanding the function $S(x)$ in powers of a small parameter $\delta$, $S(x) = \delta^{-1} \sum_{n=0}^{\infty} \delta^n S_n(x)$, and inserting it into the Helmholtz Eq. (1) to leading order, we can show that in the limit of $\delta \to 0$, $\delta$ scales with $k^{-1}$. Setting $\delta = 1/k$ and collecting terms with the same power of $k$, we can write down the two dominant terms, namely $k^2 = 1/\delta^2$, and $k^1 = 1/\delta^1$. To each of these terms a corresponding equation is found:

$$\text{Re}[\varepsilon(x)] + i\,\text{Im}[\varepsilon(x)] - [S_0'(x)]^2 = 0 \qquad (0.1)$$

$$iS_0''(x) - 2S_0'(x)S_1'(x) = 0. \qquad (0.2)$$

The exactness requirement of our ansatz necessitates that all terms $S_{n>0}$ are zero and the demand for constant intensity of $\psi(x)$ calls for a real-valued $S_0(x)$. Both conditions can be fulfilled by choosing $\text{Im}[\varepsilon(x)] = -S_0''(x)/k$ such that the term $\text{Im}[\varepsilon(x)]$ moves from Eq. (0.1) to Eq. (0.2), leading to $\text{Re}[\varepsilon(x)] = [S_0'(x)]^2$ and

$S_1'(x) = 0$. As a result $S_1(x) = const.$ and all higher terms are constant as well.

Setting $S'_0(x) = W(x)$, we finally obtain the non-Hermitian dielectric function (relative permittivity), as shown in Eq. (2).

**Iteration technique for determination of the gain-loss profile given the $n_R(x)$ distribution.** Given the function $W(x)$, the dielectric distribution $\varepsilon(x)$ and the corresponding refractive index distribution can be directly determined and vice versa. The same is, however, not true if, as a starting point, the real part of the refractive index $n_R(x)$ is known instead (typical situation in many realistic cases). The reason is that by adding an imaginary part of the refractive index $n_I(x)$ (e.g., by pumping), one not only changes $\text{Im}[\varepsilon(x)]$ but also $\text{Re}[\varepsilon(x)]$. In other words, by adding gain and loss to a material, the real part of the dielectric function changes as well. In order to overcome such a problem, we employ an iterative technique, that is based on the following expressions between $W(x)$ and the complex index of refraction: $n_R^2(x) - n_I^2(x) = W^2(x)$ and $2n_R(x)n_I(x) = -[\partial_x W(x)]/k$. To identify $n_I(x)$ and $W(x)$ required for CI-states (that correspond to a specific and given $n_R(x)$ distribution), we applied an iterative numerical scheme that allows to determine these unknown distributions, by starting with a reasonable guess function for $W(x)$. In particular, at the $m^{th}$ step of iteration the equations read:

$$n_I^{(m+1)} = -W^{'(m)}/(2kn_R) \tag{0.3}$$

$$W^{(m+1)} = [n_R^2 - n_I^{2(m+1)}]^{1/2}, \tag{0.4}$$

where $k$ is the given wavenumber of the incident plane wave. We explicitly checked that the method converges and leads to the desired results.

**Effects of gain saturation.** We prove here that the constant-intensity state can be a scattering eigenstate mode under perfect transmission boundary conditions in a medium with gain saturation[74]. The corresponding nonlinear Helmholtz equation is:

$$\frac{d^2 U}{dx^2} + k^2[\varepsilon_R(x) + i\varepsilon_L(x)]U + \frac{ik^2 P(x)}{1+\Gamma|U|^2}U = 0, \quad (0.5)$$

where $\varepsilon_L(x) > 0$ describes the material loss in the cavity, $P(x) < 0$ the pump, and $\Gamma$ the saturation coefficient. We are looking for constant-intensity states of the form

$$U(x) = A e^{ik\int W(x')dx'}, \quad (0.6)$$

where the amplitude $A$ of the solution plays an important role now. The nonlinear Eq. (0.4) can be written in the form of Eq. (1) with an effective complex permittivity $\varepsilon_{\it eff}(x)$, defined as follows:

$$\varepsilon_{\it eff}(x) = \varepsilon_R(x) + i[\varepsilon_L(x) + \frac{P(x)}{1+\Gamma|U|^2}]. \quad (0.7)$$

By substitution of Eq. (0.6) into Eq. (0.4) we obtain the following equations:

$$\mathrm{Re}[\varepsilon_{\it eff}(x)] = \varepsilon_R(x) = W^2(x), \quad (0.8)$$

$$\mathrm{Im}[\varepsilon_{\it eff}(x)] = \varepsilon_L(x) + \frac{P(x)}{1+\Gamma A^2} = -\frac{1}{k}\frac{dW}{dx}. \quad (0.9)$$

Given the auxiliary function $W(x)$ the real part of the permittivity can be easily determined by the above Eq. (0.8). Regarding the imaginary part of the permittivity, and, the pump profile $P(x)$, these can be determined as follows:

$$\varepsilon_L(x) = \begin{cases} k^{-1} dW/dx, & W' < 0, \\ 0, & W' > 0 \end{cases} \qquad (0.10)$$

$$P(x) = \begin{cases} (1+\Gamma A^2) k^{-1} dW/dx & W' > 0, \\ 0, & W' < 0. \end{cases} \qquad (0.11)$$